# Conceptual architecture of the detector infrastructure for WST

A. Meoli, O. Iwert, E. George, O. Squalli, M. Richerzhagen,
A. Rüde, M. Seidel, S. Lévêque

European Southern Observatory (ESO), Germany

## ABSTRACT

The Wide-field Spectroscopic Telescope (WST) is a proposed 12 m wide-field spectroscopic facility combining several multi-object spectrographs. It requires a yet unprecedented number of detectors. In this paper we present conceptual architecture for the detector controller and infrastructure required to operate a large number of detectors, potentially applicable to WST, focusing on the system-level, power distribution, and the associated data handling. We also consider how these elements may evolve over the expected development timeline of such a facility. Motivated by the scale of the problem, we outline a possible distributed detector-controller architecture, based on modular units placed close to the detectors and networked backend electronics.

**Keywords:** WST, detector infrastructure, distributed detector controller

## 1. INTRODUCTION

The Wide-field Spectroscopic Telescope (WST) is a proposed 12 m spectroscopic survey facility providing the simultaneous operation of a high-multiplex (>20,000-fibre) multi-object spectrograph (MOS), in low- and high-resolution variants, and an integral field spectrograph (IFS) [1, 2]. Its instrumentation is built from many nominally identical spectrograph modules [2], and the number of detectors across the three instruments is of order ~750 units. Given the detector numbers, the detector controller and its infrastructure is a facility-scale system problem. Lee et al. [2] note that the electronics and cooling for 144 IFS spectrographs "will be extremely complex and require significant volume," and flag the choice between external and integrated detector controllers as a dedicated technology study; Richerzhagen et al. [3] survey the trade space and propose, as a complement to ESO's general-purpose New General detector Controller, second generation (NGCII) [3, 4], a distributed approach for instruments built from many instances of the same detector.

Basing the architecture on NGCII, WST would approximately 2 height-unit 19" device unit per detector [3, 4], each sitting within approx. 10 m of its cryostat, with complex and expensive cables and connectors [3].

The distributed alternative replaces this with one digital Ethernet link per detector, carrying packetized pixels, power and timing on a single cable, compacting the per-detector controller into a small module at the detector and the rack-scale plant into a set of networked backend shelves. The consequence is an order-of-magnitude reduction in cable cross-section at the cryostat.

The framing is conceptual, no technology or vendor is committed and numbers bound what the architecture must accommodate.

## 2. SCALE AND REQUIREMENTS

The infrastructure is sized by three external inputs: detector population, sampling cadence from the photometric budget, and integration. All numbers below are concept stage estimates that bound what the architecture must accommodate and are treated as order of magnitude.

WST foresees three instruments sharing the focal plane [1, 2]: a panoramic integral-field spectrograph (IFS), built from of order two hundred identical 2-detector modules; a low-resolution multi-object spectrograph (MOS-LR), several tens of 4-detector modules and a smaller high-resolution MOS (MOS-HR), of order ten 4-detector modules. The designs are under active trade-off, so the population is an envelope of order ~750 detectors facility-wide across the design options under study [2, 12]. The architecture is per instrument with no cross instrument data sharing, so we sized against a single worst-case instrument, considering approximately 400 detectors.

The WST instrument trade-off retains CMOS detectors, favored for a ~60 % lower operational carbon footprint than cryocooled CCD options [12]. The science exposure is built from sub-exposures: the white paper [1] quotes a ~1 h signal-to-noise unit, split in practice into order four ~15-minute sub-exposures for cosmic-ray rejection and dithering. We take a 15-minute sub-exposure as the sizing integration, sampled non-destructively M times in the up-the-ramp (UTR) regime. The Garnett & Forrest formula [5]

$$\sigma^2(Q) = \sigma_{read}^2 \cdot 12(M-1)/[M(M+1)]$$

gives the read-noise contribution to the integrated signal Q; for large M, $\sigma(Q) \approx \sigma_{read} \cdot \sqrt{(12/M)}$. The target is an effective read noise on the integrated ramp of $\sigma(Q) < 1\ e^-$.
We evaluate four single-read RON scenarios, from deliberately demanding cases:

Table 1. $M_{min}$ and required cadence to reach $\sigma(Q) < 1\ e^-$ for the four RON scenarios.

| **$\sigma_{read}$** | **Mmin for $\sigma(Q) < 1\ e^-$** | **Required cadence in 900 s** |
|---|---|---|
| 3 $e^-$ | 106 | 0.12 fps |
| 5 $e^-$ | 298 | 0.33 fps |
| 10 $e^-$ | 1198 | 1.33 fps |
| 15 $e^-$ | 2698 | 3.00 fps |

This scaling is deliberately optimistic since it assumes read noise that is white and uncorrelated from sample to sample, so that M averages it down indefinitely. In a real CMOS sensor several contributions break this assumption, 1/f and other low frequency noises sets a floor once the ramp duration exceeds the inverse corner frequency, after which the effective noise no longer integrates down as $\sqrt{(12/M)}$, correlated row and common-mode noise are only partially removable by reference-pixel correction and dark-current shot noise and image lag accumulate over the ramp independently of M.
The $M_{min}$ values of Table 1 are therefore lower bounds, computed for sizing the transport and compute envelopes.
A dedicated assessment is planned as next step of this study: end-to-end modelling of sensors noise with Pyxel, the open-source detector simulation framework developed by ESA and ESO [14] to derive detector trade-off requirements.
We adopt 1 frame/s as the architectural ceiling: a class needing a higher cadence to close the budget in 15 minutes falls outside the feasibility envelope. The target class is the cooled scientific CMOS envelope $\sigma_read \approx 3\text{-}5e^-$ ($M \approx 106\text{-}298$), well below the ceiling and with headroom for the extra UTR samples used by the cosmic-ray rejection.
The cadence span produces a span of raw bandwidths at each detector's output:

Table 2. Bandwidth envelope, 16-bit pixels. The aggregate column is the worst-case sizing instrument.

| Regime | M | Per-detector 6k | Per-detector 12k | Instrument |
|---|---|---|---|---|
| Nominal, $\sigma_{read} = 3\ e^-$ | 106 | 71 Mbit/s | 285 Mbit/s | 28 Gbit/s |
| Nominal, $\sigma_{read} = 5\ e^-$ | 298 | 200 Mbit/s | 800 Mbit/s | 80 Gbit/s |
| Ceiling at 1 fps cap | 900 | 604 Mbit/s | 2.4 Gbit/s | 242 Gbit/s |

In the nominal regime, a 1 GbE link is sufficient for the 6k detector scenario, still providing about 5× margin. For the 12k scenario, instead, the link approaches saturation at the upper end of the operating range, so a 2.5 GbE interface may be more appropriate.

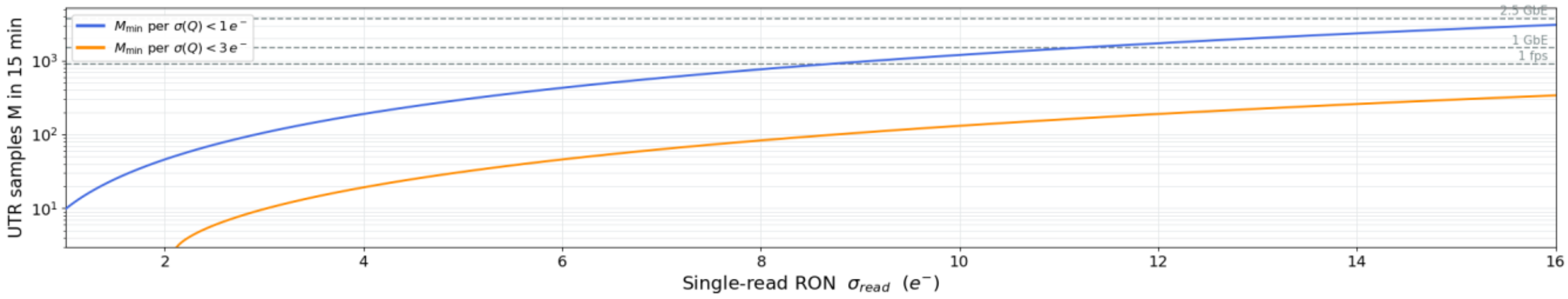

Figure 1. Minimum number of UTR samples required within a 15 min integration to reach an effective noise of $1e^-$ (blue) and $<3e^-$(orange), as a function of the single-read noise. The horizontal dashed lines give the maximum M supported by the frame cadence; the intersection with each curve defines the largest single-read noise compatible the sampling rate the processing node must sustain.

At the maximum frame rate of 1 fps, the 6k detector requires 2.5 GbE, while the 12k detector requires 10 GbE. At instrument level, the total data rate ranges from approximately 28 Gbit/s to about 80 Gbit/s meaning that a single 100 GbE link would remain sufficient.

## 3. DISTRIBUTED DETECTOR-CONTROLLER ARCHITECTURE

The purpose of the proximal module is to collect the parallel sensor data, perform the required data shaping close to the detector, and output a single Ethernet stream matched to the actual science acquisition rate.
The architecture is organized around a transport layer that moves pixels from each detector to the processing, and a processing layer that performs the first reduction steps. Physically, the transport layer terminates at an edge aggregator, where the processing begins and from which data propagates to the facility data system. One layer can be reimplemented independently of the other, assuming that packet format, timing semantics, QoS, etc., are preserved.
The detector proximity module is the minimal unit placed close to the detector. Its role is to serialize and packetize the pixel stream, timestamp each frame against the facility timing distribution, buffer frames in local DDR-SDRAM and forward them on a standard Ethernet output. Each module serves one detector. Crucially, the module is warm electronics close to the detector, not cold electronics inside the cryostat: digitizing at the detector removes the long analog runs of a centralized warm controller, while staying warm avoids the principal obstacle to cold readout, as commercial components are specified only to −40 °C, and at the −55 °C military range key parts such as DDR-SDRAM and >1 Gbit/s Ethernet PHYs may be unavailable [3].
The detector module is intentionally minimal because it must be produced in volume, qualified once, and replaced only on failure.

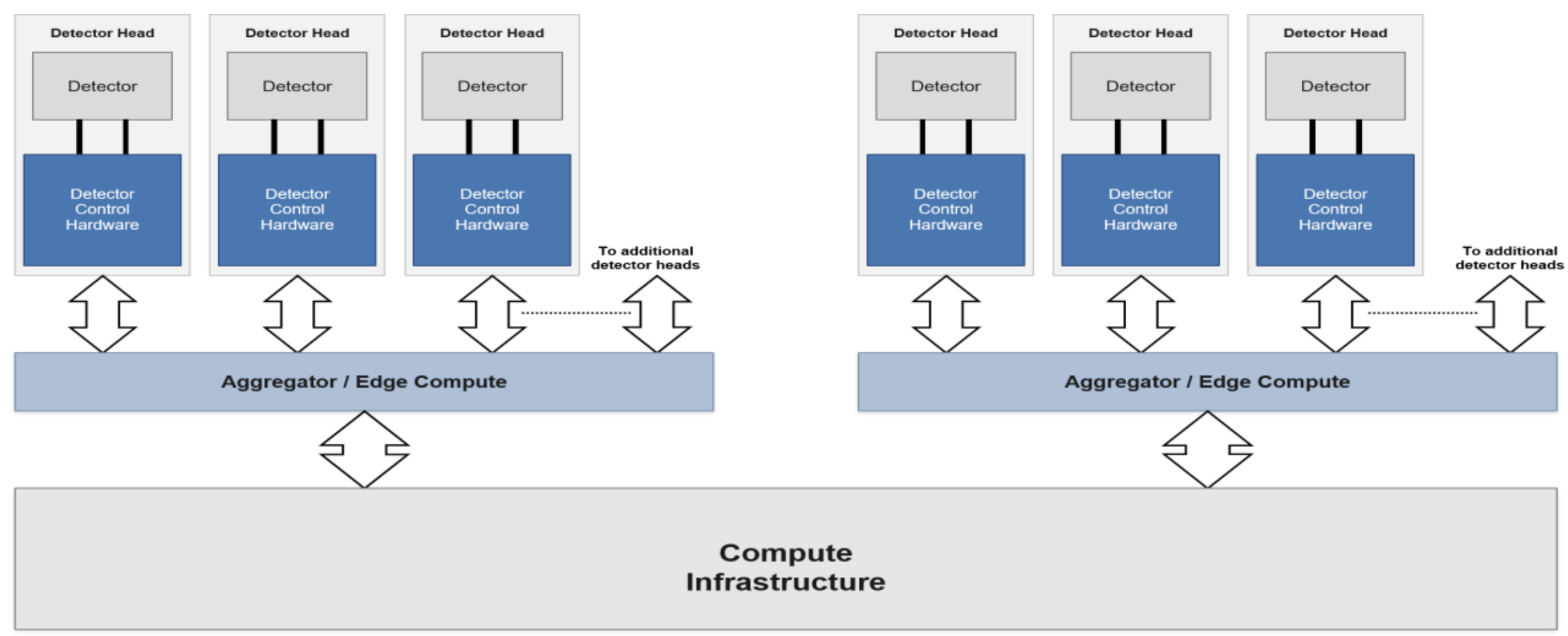

Figure 1. Proposed architecture for the distributed detector readout. Each detector module couples a detector to its local control hardware; groups of detector modules are served by an aggregator / edge-compute node, which performs the first data reduction before forwarding to the compute infrastructure.

The edge aggregator is the networked backend unit: it terminates up to 32 links per shelf and executes the in-line UTR processing. The aggregation is done within each instrument, giving per-instrument partitioning with no cross-instrument data sharing. We estimated approximately 25 2U chassis for the worst-case instrument. The aggregator shelf is required to be mature and hot-swappable, so xTCA-family carriers [10] are good candidates, given also the NGCII heritage.
This architecture adapts the high-energy-physics data-acquisition paradigm [6, 7] to astronomical instrumentation.

## 4. POWER DISTRIBUTION

Every detector is powered over the same Cat6A that carries its data and timing, exploiting the ability of copper Ethernet to embed synchronization and power in the same link with PoE.
The per-unit budget is dominated by the proximal electronics: a fully-digital 6k detector dissipates less than 2 W [3], and the proximity module (FPGA, DDR-SDRAM, Ethernet PHY/PoE front-end, bias generation) brings the total up to 20 W, consistent with the ~15 W per-unit FPGA readout assumed in the WST life-cycle assessment [12]; the 12k tier, with a larger sensor and a 10 GbE PHY, could reach up to 45 W. These budgets set the PoE class per scenario:

- PoE type 3 for the 6k scenario
- PoE type 4 for the 12k scenario

Across the sizing instrument, 25 shelves would consume up to 15 kW and over the facility in the order of 30 kW.
A modest number at facility scale, where detector cooling dominates energy use [2].
These figures are concept-stage allocations, to be consolidated against the selected sensor and FPGA classes at preliminary design.

## 5. DATA HANDLING

### *5.1 Continuous UTR streaming*

A packetized readout of many detectors where there are extreme data bursts can overwhelm switch buffers and cause packet loss unless mitigated by QoS strategies and coordinated traffic shaping to prevent simultaneous bursts from saturating the links [3].
Under UTR sampling this congestion is removed by construction, because the time constraint is on the capture of each frame. The detector module buffers each frame in DDR-SDRAM and spreads its transmission over the full inter-read interval.
A burst regime peaks at ~16 Gbit/s, while continuous streaming holds the aggregate at the shelf's sustained input rate, ~2 Gbit/s at this cadence.
The same DDR act as a retransmission buffer: every frame remains locally available for the full interval, so a lightweight retransmit check makes the transport lossless at no extra hardware. A concept-level controller should operate in the continuous-streaming regime by design.

### *5.2 Edge reduction and data products*

The processing plane performs the first reduction of the UTR sequence into data products.
A per-pixel slope fit and variance (Garnett & Forrest [5] weighting) implemented as running sums of x, y, $x^2$, xy, each frame updates the per-pixel state in O(1) without retaining the ramp, at of order ~4 FLOP per input byte.
Cosmic-ray rejection runs as streaming jump detection on consecutive differences, along the lines of established UTR ramp processing [13]: each new read forms a difference $d_i = s_i - s_{i-1}$, standardized against the running rate estimate and its variance; an outlier marks a jump, the fit is segmented at the hit, and slope, variance and CR count are written out.
As for the ramp fitting, the algorithm requires only (O(1)) state per pixel. Storing the full ramp would not be necessary or realistically feasible. The resulting computational intensity is of the order of 6 FLOP/byte.
Both kernels assume white frame to frame residuals, the Pyxel model could be used to generate synthetic ramps with correlated noise to quantify the resulting slope bias and CR false-alarm rate.
Combining the two, the arithmetic intensity of edge reduction is ~10 FLOP/byte; applied to the per-aggregator throughput this is ~1.4–4.0 GFLOP/s per aggregator in the nominal regime (25 chassis), rising to ~12 GFLOP/s per aggregator (~300 GFLOP/s instrument, ~0.7 TFLOP/s facility) at the ceiling.
At an arithmetic intensity of order 10 FLOP/byte, the reduction sits below the ridge point of a Xeon D-class CPU and is therefore bandwidth limited. Even in the most demanding operating condition the required throughput stays one to two orders of magnitude below the platform capability.
On an embedded CPU, the reduction is therefore limited mainly by memory bandwidth rather than by computing capability. Even in the most demanding operating condition, the required throughput remains one to two orders of magnitude below the expected platform capability.
The algorithms considered here serve as an order of magnitude check on the sizing of the processing requirements rather, additional processing steps would increase the arithmetic. A GPU module is therefore not a strict requirement for the baseline implementation, but a dedicated slot is retained as a provision for future extensions, introducible without altering the system architecture.
This local reduction is also essential to keep the archived data volume within the few PB per year range considered for WST data handling. Since storage is expected to dominate the 20-year life-cycle footprint [12], the edge-processing stage contributes directly to the sustainability of the facility.

### *5.3 Timing*

The inter-frame interval is in the order of 1 to 10 seconds, and millisecond accuracy is sufficient for UTR timestamping.

The architecture could distribute a single facility-wide PTPv2 grandmaster [9] through PTP-capable industrial Ethernet switches, with boundary clocks at each instrument-aggregator domain isolating one instrument's synchronization timescale from the others. PTPv2 is over-specified by orders of magnitude here, but comes essentially for free from the switches already in the transport plane and absorbs any future cadence tightening; White Rabbit [11] would add another order of magnitude at the cost of a dedicated subsystem, unjustified on this envelope.

## 6. CONCLUSIONS

The detector population foreseen for WST goes clearly beyond the practical scaling limit of a centralized controller architecture. An NGCII-class implementation would require approximately one 2U controller unit per detector [3, 4], which becomes difficult to justify when extended to the full instrument population.
Starting from the distributed approach proposed by Richerzhagen et al. [3], we have developed a WST-specific conceptual architecture based on two levels: a minimal warm detector module located close to each detector, and a set of xTCA aggregators allocated per instrument. At this stage, the internal implementation is intentionally left open, while only the interfaces between the two levels are frozen.
Continuous UTR streaming removes the simultaneous-readout burst by construction. Across the nominal operating regime, this reduces the aggregate peak traffic by approximately a factor of 5 to 14, without requiring additional hardware complexity.
The same Cat6A cable is used for detector data, timing and power delivery. PoE Type 3 is sufficient for the 6k scenario, while PoE Type 4 is so for the 12k scenario. In this way, a common cabling concept can be maintained over the complete detector population.
Approximately 60 aggregators are partitioned by instrument, with no cross-instrument sharing of detector data. This has an important consequence: the 6k scenario required by IFS and MOS-LR can proceed independently, while the 12k detector and FPGA decision can be postponed until the MOS-HR development reaches sufficient maturity.
The present sizing is based on an intentionally optimistic white-noise UTR model. This assumption is useful for establishing the first-order architecture, but it is not sufficient to close the detector requirements. The immediate continuation of the study is therefore a dedicated detector assessment, combining Pyxel end-to-end simulations [14] with laboratory characterization. The objective is to derive quantitative trade-off requirements among the read-noise spectrum, the sampling cadence and the sub-electron performance target.

## ACKNOWLEDGEMENTS

The authors acknowledge the WST consortium, ESO and the partner institutes. This work is part of ESO's contribution to the Detector Work Package WP4.7 of the "Research and Innovation Action funded by the European Research Executive Agency (REA) under the powers delegated by the European Commission under Grant Agreement No. 101183153 - WST - HORIZON-INFRA-2024-DEV-01"